\documentclass[lettersize,journal]{IEEEtran}
\usepackage{amsmath,amssymb,amsfonts}
\usepackage[caption=false,font=normalsize]{subfig}
\usepackage[dvipdfmx]{graphicx}
\usepackage{cite}
\usepackage[hidelinks, bookmarks=true,
bookmarksnumbered=true, bookmarkstype=toc,
colorlinks=true, allcolors=blue]{hyperref}
\usepackage{siunitx}
\usepackage{multirow}
\usepackage{threeparttable}
\graphicspath{{./figures/}}
\DeclareGraphicsExtensions{.pdf,.jpg,.png}

\begin{document}
\title{Design Method of Quasi-Lumped Element Bandpass Filters Using Superconducting
Coplanar Waveguide for Millimeter-Wave Multichroic Imaging}

\author{Shinsuke~Uno,
        Kah~Wuy~Chin,
        Tai~Oshima,
        Satoshi~Ono,
        Takeshi~Sakai,
        Kazuki~Watanabe,
        Shuhei~Inoue,
        Tatsuya~Takekoshi,
        and~Kotaro~Kohno%
\thanks{Manuscript received 29 July 2025; revised 25 May 2026; accepted 24 June 2026.
Date of publication XXX XXX 2026; date of current version 7 July 2026.
{\it (Corresponding author: Shinsuke Uno.)}}
\thanks{Shinsuke Uno is with RIKEN Center for Advanced Photonics, Saitama 351-0198, Japan (email: shinsuke.uno@riken.jp).}
\thanks{Kah Wuy Chin was with the University of Tokyo, Tokyo 181-8588, Japan.}
\thanks{Tai Oshima and Kazuki Watanabe are with National Astronomical Observatory of Japan (NAOJ), Tokyo 181-8588, Japan, 
and also with the Graduate University for Advanced Studies (SOKENDAI), Tokyo 181-8588, Japan.}
\thanks{Satoshi Ono and Takeshi Sakai are with the University of Electro-Communications, Tokyo 182-8585, Japan.}
\thanks{Shuhei Inoue and Kotaro Kohno are with the University of Tokyo, Tokyo 181-8588, Japan.}
\thanks{Tatsuya Takekoshi is with Kitami Institute of Technology, Hokkaido 090-8507, Japan.}
}

\markboth{IEEE Transactions on Applied Superconductivity,~Vol.~XXX, No.~XXX, XXX~2026}%
{Uno \MakeLowercase{\textit{et al.}}: XXX Title}

\IEEEpubid{%
\raisebox{-5mm}{%
\makebox[\linewidth]{%
\fbox{%
\parbox{0.85\textwidth}{\footnotesize
\copyright~2026 IEEE. Personal use of this material is permitted. Permission from IEEE must be obtained for all other uses, in any current or future media, including reprinting/republishing this material for advertising or promotional purposes, creating new collective works, for resale or redistribution to servers or lists, or reuse of any copyrighted component of this work in other works.
}}}%
\hfill}
}

\maketitle

\begin{abstract}
An on-chip band-defining filter coupled with a superconducting photon detector is a promising technology for developing multi-band imaging cameras at millimeter and submillimeter wavelengths.
In this paper, we present the design of on-chip bandpass filters based on coplanar waveguide geometry, which can be easily integrated into large-format multi-band detector arrays.
A lumped element filter design is suitable not only for achieving a compact footprint but also for suppressing harmonics to reduce band-to-band crosstalk in a multiplexer.
However, the coplanar waveguide geometry and the photolithography process rule limit the maximum available inductance and capacitance of lumped elements, which does not sufficiently meet the requirements of filter circuits.
To overcome this limitation, we have established a design method for quasi-lumped element filters, in which the maximum element size is relaxed to a quarter wavelength, exceeding the ideal lumped element size.
We achieved design solutions for 150, 220, and 270~GHz 8th-order Chebyshev bandpass filters and a triplexer.
We also report on the measurement results of a scaled model of the bandpass filter, demonstrating the validity of our proposed filter design.

\end{abstract}

\begin{IEEEkeywords}
Millimeter/submillimeter astronomy, on-chip filters, coplanar waveguide, superconducting circuit devices.
\end{IEEEkeywords}

\section{Introduction}
\IEEEPARstart{S}{uperconducting} integrated circuit technology is of great interest for instrumentation in millimeter and submillimeter astronomy.
There have been large-format arrays of pixels, consisting of wideband antennae, multiple band-defining filters, and superconducting detectors, which have been developed to enable multi-band imaging observations and cosmic microwave background experiments \cite{Sayers2014_MUSIC,Thornton2016_ACTPol,Westbrook2018_PB2,Sobrin2022_SPT-3G}.
While multi-band imaging is also realized by splitting bands with quasi-optical filters \cite{Wilson2020_TolTEC}, this approach suffers from the demand for huge cold optics as the number of bands increases.
On-chip filters provide a compact and scalable solution for multi-band, wide-field imaging observations.
\IEEEpubidadjcol

The microstrip line structure has generally been used in on-chip filters \cite{Kumar2009,Suzuki2012,thesis_OBrient2010,thesis_Suzuki2013}.
A microstrip line consists of strip and ground layers separated by a thin dielectric, offering high design flexibility.
However, microstrip-based bandpass filters require photolithography on both the strip and the ground layers to realize large series capacitances for lumped element filters, or require vias from the strip to the ground for distributed stub filters.
The multilayer configuration of microstrip lines makes it difficult to develop a uniform pixel array on a wafer.
Microstrip lines also suffer from dielectric loss depending on the dielectric material used, leading to a reduction of the optical throughput.
In this work, we chose the coplanar waveguide (CPW) design as an alternative solution for on-chip filters.
CPW features strip and ground planes formed in a single metal layer on a substrate.
Because of its simple fabrication process with fewer parameters compared with microstrip lines, the CPW architecture is easily developed into large pixel arrays.
For example, 350~GHz bandpass filters based on CPW stepped impedance resonators were demonstrated \cite{Ding2021}.

In this paper, we present a design method for CPW-based quasi-lumped element bandpass filters (BPFs).
The theoretical details are also described in \cite{thesis_Uno2024}.
We start from lumped element BPF circuits rather than resonators, aiming for a compact footprint and effective rejection of higher-order harmonics towards further multiplexing.
As discussed below, the BPF circuit will be optimized as a quasi-lumped element due to the limited design flexibility of single-layer CPW.
We designed BPFs with center frequencies of 150, 220, and 270~GHz, each with a bandwidth of 40--45~GHz to match the atmospheric window.
Each BPF is designed as an 8th-order Chebyshev filter to achieve a wide fractional bandwidth of 0.15--0.3 and suppress band-to-band crosstalk when used in a triplexer.
We also measured the frequency response of a scaled model of a BPF at 7.5~GHz to demonstrate the validity of our quasi-lumped element BPF design.

\color{red}
\begin{figure*}[tb]
        \centering
        \includegraphics[width=\linewidth]{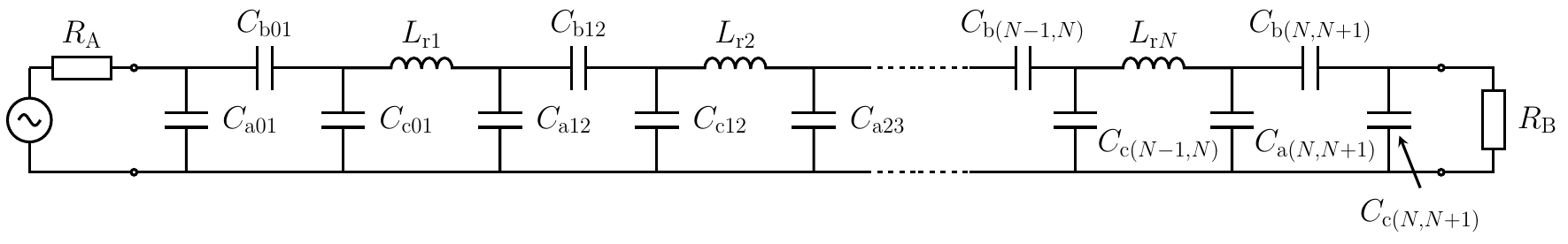}
        \caption{Equivalent lumped element circuit of an $N$th-order BPF.
        The transformation follows the method of Nomura \& Kobayashi (1996) \cite{Nomura1996}.}
        \label{fig:circ_bpf}
\end{figure*}

\begin{table*}[tb]
    \caption{Ideal lumped element values for the 8th-order BPFs}
    \label{tab:lebpf_params}
    \centering
    \begin{tabular}{ccccccccccccccccc} \hline
         $f_0$ & $BW$ & $L_\mathrm{r}$ & $C_{\mathrm{a}01}$ & $C_{\mathrm{b}01}$ & $C_{\mathrm{c}01}$ & $C_{\mathrm{a}12}$ & $C_{\mathrm{b}12}$ & $C_{\mathrm{c}12}$ & $C_{\mathrm{a}23}$ & $C_{\mathrm{b}23}$ & $C_{\mathrm{c}23}$ & $C_{\mathrm{a}34}$ & $C_{\mathrm{b}34}$ & $C_{\mathrm{c}34}$ & $C_{\mathrm{a}45}$ & $C_{\mathrm{b}45}$ \\
         (GHz) & (GHz) & (pH) & (fF) & (fF) & (fF) & (fF) & (fF) & (fF) & (fF) & (fF) & (fF) & (fF) & (fF) & (fF) & (fF) & (fF) \\\hline
        150 & 45 & \textbf{50.3} & 4.9 & 44.1 & 22.6 & \textbf{62.6} & \textbf{44.1} & 16.8 & 35.5 & 17.8 & 32.6 & 34.2 & 16.1 & 33.7 & 34.0 & 15.9 \\
        220 & 40 & \textbf{46.7} & 3.2 & 9.5 & 11.7 & \textbf{27.9} & \textbf{9.5} & 14.7 & 19.2 & 4.9 & 18.4 & 18.9 & 4.5 & 18.7 & 18.8 & 4.5 \\
        270 & 40 & \textbf{46.7} & 2.0 & 4.9 & 7.8 & \textbf{18.8} & \textbf{5.0} & 10.6 & 13.1 & 2.6 & 12.7 & 12.9 & 2.4 & 12.8 & 12.9 & 2.4 \\\hline
    \end{tabular}
\end{table*}
\color{black}

\section{Design of Quasi-Lumped Element BPFs}
To construct a filter as a lumped element circuit, each element should be small enough relative to the wavelength.
Otherwise, parasitic components become non-negligible and may create spurious passbands.
We defined the lumped element criterion that the element size is smaller than $1/8$ of the guided wavelength, and assessed the feasibility of lumped element solutions.
For 150, 220, and 270~GHz BPFs, we assume the maximum operational frequency of 350~GHz, corresponding to a lumped element size of $\lesssim$ 35~$\mu$m.

\subsection{Lumped Element Equivalent Circuit Design}
\label{subsec:theory}
A lumped element BPF equivalent circuit is constructed based on the network synthesis method \cite{Matthaei1980,Pozar2011}, and transformation proposed by Nomura \& Kobayashi (1996) \cite{Nomura1996}.
The equivalent circuit of an $N$th-order BPF is shown in Fig.~\ref{fig:circ_bpf}.
In this circuit, $\pi$-networks of capacitors $C_{\mathrm{a}(n,n+1)}$, $C_{\mathrm{b}(n,n+1)}$, $C_{\mathrm{c}(n,n+1)}$ are cascaded with series inductors $L_{\mathrm{r}n}$.
Note that this circuit design has impedance inverters at the input and output stages, functioning as impedance matching networks to the CPW transmission line.
This is different from conventional lumped element filter circuits based on microstrip lines \cite{Kumar2009,Suzuki2012,thesis_OBrient2010,thesis_Suzuki2013} and is advantageous for realizing BPFs using CPW structures with limited series inductance and capacitance.

We have set 8th-order Chebyshev BPFs with three bands of center frequency ($f_0$) and bandwidth ($BW$).
Here we assumed that at each filter a ripple level is 0.2~dB and port impedances ($R_\mathrm{A}, R_\mathrm{B}$) were both 90~$\Omega$.
The derived lumped element values are listed in Table~\ref{tab:lebpf_params}.
Here we can treat $L_{\mathrm{r}n}$ as free parameters.
In practice, we used a common value $L_{\mathrm{r}n} = L_\mathrm{r} \ (n=1, \dots, N)$ for the series inductors for simplicity.
Regarding the symmetry, $C_{\mathrm{a}(n,n+1)} = C_{\mathrm{c}(N-n,N-n+1)}$ and $C_{\mathrm{b}(n,n+1)} = C_{\mathrm{b}(N-n,N-n+1)} \ (n=0, \dots, N)$ are applied.
To minimize the required shunt and series capacitances, we gave $L_{\mathrm{r}}$ in Table~\ref{tab:lebpf_params} with the maximum value allowed under the lumped element criterion of a 35~$\mu$m length, as detailed in the following section.

\subsection{Filter Layout and Element Design}
\begin{figure*}[!t]
        \centering
        \includegraphics[width=\linewidth]{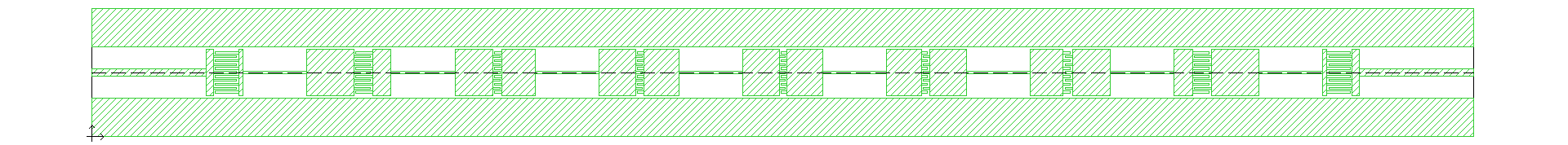}
        \caption{Layout of the 8th-order 150~GHz quasi-lumped element BPF in Sonnet, reproduced from \cite{thesis_Uno2024}.
        The superconducting film and substrate are shown in green and blank, respectively.
        The central dashed line indicates the symmetry axis.}
        \label{fig:layout_bpf}
\end{figure*}

The filter layout is shown in Fig.~\ref{fig:layout_bpf}, reproduced from \cite{thesis_Uno2024}.
The series inductors and shunt capacitors are represented by finite CPW lengths with narrow and wide strips, respectively.
Series capacitors are realized by strip gaps or interdigitated capacitors (IDCs).
The minimum pattern size is set to 2~$\mu$m, limited by our photolithography process.
Since the radiation loss increases as the CPW width increases, we limit the width to 52~$\mu$m for 150~GHz BPF and to 28~$\mu$m for 220~GHz and 270~GHz BPFs.

The filter layer is a superconducting film deposited on the silicon substrate.
Throughout the BPF design, we assume a zero-thickness plane with a sheet inductance of $L_\mathrm{s}=$1~pH$/\square$, representing a superconducting NbTiN film.
The silicon substrate is 200~$\mu$m thick with relative permittivity $\varepsilon_\mathrm{r}=$ 11.45 and no dielectric loss.
Given close contact with a metal holder, there is an additional perfect electric conductor (PEC) plane on the backside of the substrate.
The frequency response of each circuit element is simulated using Sonnet \cite{Sonnet_webpage}.
Note that our simulation model in Sonnet omits radiation loss in order to reduce computational cost during the design and optimization.

\subsubsection{Series inductor and shunt capacitor}
\begin{figure}[tb]
        \centering
        \subfloat[]{\includegraphics[scale=0.6]{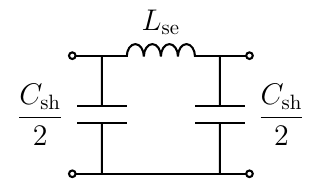}
        \label{subfig:circ_tl}}
        \hfil
        \subfloat[]{\includegraphics[scale=0.6]{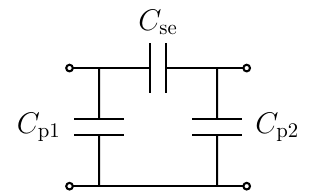}
        \label{subfig:circ_gapc}}
        \caption{Equivalent lumped element circuits. 
        (a) Short transmission line for a series inductor or a shunt capacitor. 
        (b) Gap and IDC for a series capacitor.}
        \label{fig:equiv_circ}
\end{figure}

A short CPW line with an appropriate characteristic impedance can be used as a series inductor or a shunt capacitor.
The series inductance $L_\mathrm{se}$ and shunt capacitance $C_\mathrm{sh}$ of a transmission line with physical length $l$ much shorter than the wavelength are given by:
\begin{gather}
        L_\mathrm{se} = \frac{Z_0\sqrt{\varepsilon_\mathrm{eff}}\,l}{c}, \label{eq:TL_L}\\
        C_\mathrm{sh} = \frac{\sqrt{\varepsilon_\mathrm{eff}}\,l}{Z_0c} \label{eq:TL_C},
\end{gather}
where $c$ is the speed of light, and $Z_0$ and $\varepsilon_\mathrm{eff}$ are the characteristic impedance and effective relative permittivity of the transmission line, respectively.
The equivalent $\pi$-network circuit is shown in Fig.~\ref{fig:equiv_circ}(a).
The geometrical parameters of CPW are the center strip width $w$ and gap $s$ between the strip and grounds.
For the series inductor, a high $Z_0$, narrow-strip CPW line is used.
Such a simple CPW inductor is preferred over other structures like a meander inductor, because it is less affected by parasitic components.
For the shunt capacitor, conversely, a low $Z_0$, wide-strip CPW line is used.
The properties of the CPW for series inductor and the shunt capacitor are summarized in Table \ref{tab:CPW_LE}.

\begin{table}[tb]
        \caption{Properties of the CPW transmission lines for series inductors and shunt capacitors
        (Assuming $L_\mathrm{s}=$ 1~$\mathrm{pH}/\square$)}
        \label{tab:CPW_LE}
        \centering
        \begin{tabular}{ccccccccc} \hline
                $w+2s$ & $w$ & $s$ & $Z_0$ & \multirow{2}{*}{$\varepsilon_\mathrm{eff}$} & $L_\mathrm{se}/l$ & $C_\mathrm{sh}/l$ \\
                ($\mu$m) & ($\mu$m) & ($\mu$m) & $(\Omega)$ & & (pH$/\mu$m) & (fF$/\mu$m) \\\hline
                52 & 2 & 25 & 137.0 & 9.9 & 1.44 & 0.077 \\
                52 & 48 & 2 & 28.6 & 8.3 & 0.28 & 0.336 \\
                28 & 2 & 13 & 123.4 & 10.5 & 1.33 & 0.088 \\
                28 & 24 & 2 & 34.8 & 8.8 & 0.34 & 0.284 \\\hline
        \end{tabular}
\end{table}

\subsubsection{Series capacitor}
A series gap in the center strip of CPW is used as a lumped element series capacitor.
Furthermore, an IDC with a small gap and narrow fingers is also used to obtain a relatively large series capacitance by a single layer.
These gaps or IDCs are modeled by a $\pi$-network of capacitors \cite{Simons2001}, as shown in Fig.~\ref{fig:equiv_circ}(b).
It consists of the main series capacitance and parasitic shunt capacitances between the center strip and grounds.
The series capacitance of an IDC increases with both the number and the length of its fingers.
For BPFs with width $w+2s$ of 52~$\mu$m and 28~$\mu$m, the number of IDC fingers is 11 and 5, respectively, with 2~$\mu$m line and space.
The resulting series capacitances of the 35~$\mu$m-long IDCs are about 21 and 9~fF for $w+2s$ of 52~$\mu$m and 28~$\mu$m, respectively.

\subsection{Feasibility of Lumped Element Solutions}
From the above discussion and Table~\ref{tab:CPW_LE}, the maximum available values of series inductor, shunt capacitor, and series capacitor under the lumped element criterion are summarized in Table~\ref{tab:LE_limit}.
Comparing Table~\ref{tab:lebpf_params} and Table~\ref{tab:LE_limit}, we found that all three BPFs require shunt capacitances that exceed the values available from 35~$\mu$m-long CPW lines.
In addition, these shunt capacitances cannot be adequately compensated even with the parasitic capacitances of the series inductors (CPW lines, Fig.~\ref{fig:equiv_circ}(a)) and series capacitors (IDCs, Fig.~\ref{fig:equiv_circ}(b)).
In particular, for 150~GHz BPF, the large fractional bandwidth results in a required series capacitance that is too large to be realized even with a 52~$\mu$m-wide IDC.
In conclusion, under the constraint of our 2~$\mu$m process rule, the CPW structure is unable to meet the requirements for lumped element BPF design.

\begin{table}[tb]
        \caption{Maximum available element values under the 35~$\mu\mathrm{m}$ length limit}
        \label{tab:LE_limit}
        \centering
        \begin{tabular}{ccccc} \hline
                 & $w+2s$ & max $L_\mathrm{se}$ & max $C_\mathrm{sh}$ & max $C_\mathrm{se}$ \\
                 & ($\mu$m) & (pH) & (fF) & (fF) \\\hline
                150~GHz & 52 & 50.3 & 11.8 & $\sim$ 21 \\
                220/270~GHz & 28 & 46.7 & 10.0 & $\sim$ 9 \\\hline
        \end{tabular}
\end{table}

\subsection{Solution as a Quasi-Lumped Element Filter}
To address this limitation, we developed a method to synthesize a BPF as a quasi-lumped element filter rather than an ideal lumped element filter.
By relaxing the constraint of the maximum element size from $1/8$ to a quarter of the guided wavelength, we can obtain larger inductance and capacitance with CPW structures.
A major challenge is that the substitution of quasi-lumped element structures into the equivalent circuit is not a straightforward task due to the nonlinear growth of parasitic components with increasing length.
We therefore optimized the BPF design parameters to achieve the desired bandpass response, following the steps below.

\begin{figure}[tb]
        \centering
        \includegraphics[width=\linewidth]{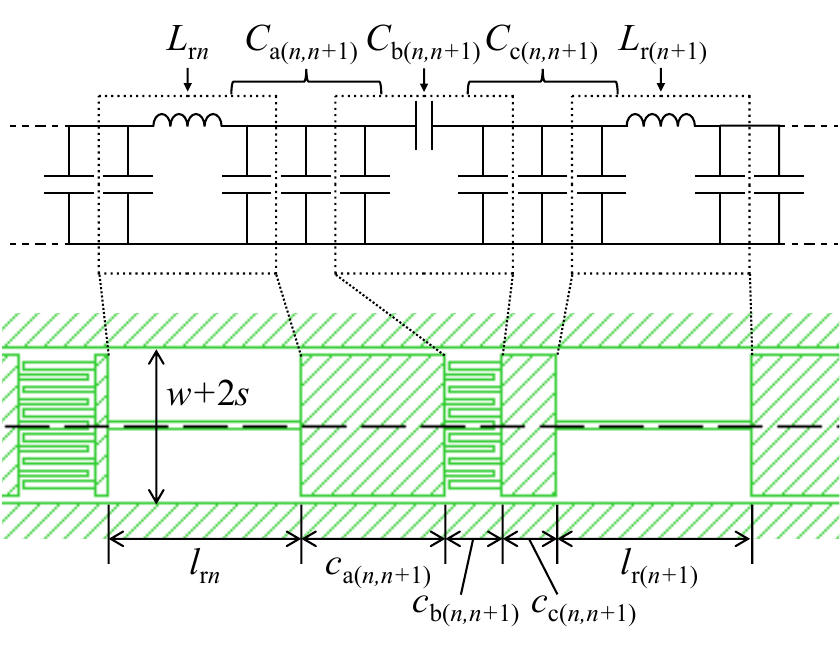}
        \caption{Sketch of the substitution of quasi-lumped element structures into the equivalent circuit.}
        \label{fig:substitute_bpf}
\end{figure}

\begin{enumerate}
        \item Initial filter parameters (i.e., center frequency, bandwidth, filter order, ripple level, and port impedances) are provided as input to determine the lumped element values.
        \item Substitution of quasi-lumped element structures to compensate parasitic shunt capacitances (Fig.~\ref{fig:substitute_bpf}).
        The series inductor is linearly scaled in length, along with the parasitic shunt capacitance, using (\ref{eq:TL_L}) and (\ref{eq:TL_C}), beyond the lumped element size limit and up to a quarter of the wavelength.
        The series capacitor is replaced by a pre-calculated scattering matrix of gap or IDC.
        After accounting for these parasitic shunt capacitances in series inductors and series capacitors, the lengths of the shunt capacitors are given by (\ref{eq:TL_C}) while ignoring parasitic series inductance.
        We also compensate for the effective electrical lengths introduced by CPW strip width discontinuities by shortening the transmission lines (order of a few $\mu$m).
        Then the lengths of all elements are determined ($l_{\mathrm{r}n}, c_{\mathrm{a}(n,n+1)}, c_{\mathrm{b}(n,n+1)}, c_{\mathrm{c}(n,n+1)}$ in Fig.~\ref{fig:substitute_bpf}).
        \item The frequency response of the BPF is calculated by cascading the scattering matrices of all the member elements.
        It was found that the realized center frequency and bandwidth exhibit systematic shifts from the input values.
        This shift is likely attributed to extra parasitic components, which are ignored in the above substitution and difficult to compensate for.
        Nevertheless, the bandpass shape is serendipitously preserved without significant distortion.
        Therefore, we can iteratively find the input values to match the realized center frequency and bandwidth as targeted.
        \item Fine-tuning is then applied to optimize the bandpass response by perturbing the design parameters.
        Using the final design parameters, the overall filter model is simulated in Sonnet to confirm the frequency response.
\end{enumerate}

In this design optimization flow, scattering matrices of individual quasi-lumped element structures are generated by fitting and interpolation to the simulation results of Sonnet.
An exhaustive parameter search can then be performed simply by cascading scattering matrices, significantly reducing computational cost as compared with full electromagnetic simulations.
This approach allows us to efficiently optimize even high-order filters with many parameters, achieving better bandpass performance and reduced crosstalk towards multiplexing.

\section{Design Results and Simulations}
\subsection{Optimal designs for 150/220/270~GHz BPFs}
\begin{figure}[tb]
        \centering
        \includegraphics[width=\linewidth]{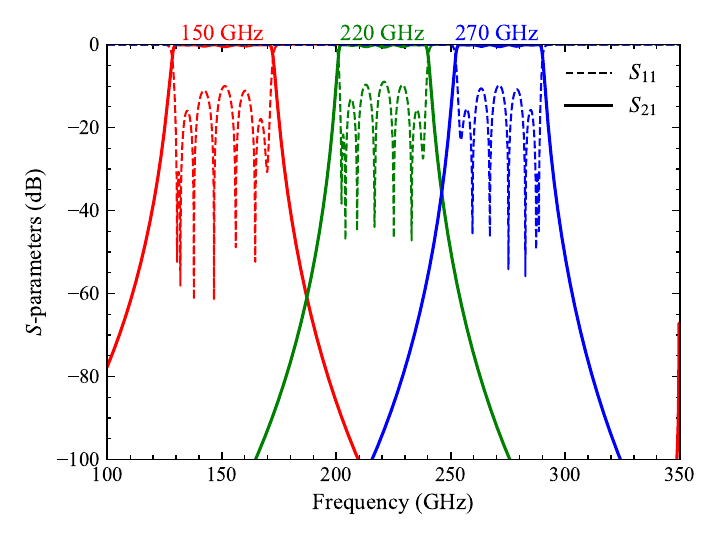}
        \caption{Simulated frequency responses of the 150, 220, and 270~GHz BPFs.}
        \label{fig:bpf_3bands_dB}
\end{figure}

We obtained quasi-lumped element design solutions for 8th-order BPFs at 150, 220, and 270~GHz.
Each BPF was designed assuming a NbTiN film of $L_\mathrm{s}=$ 1~pH$/\square$ and a port impedance of 90~$\Omega$.
Their simulated responses are shown in Fig.~\ref{fig:bpf_3bands_dB}.
In this simulation, radiation loss was again neglected in the Sonnet model.
The in-band return losses are greater than 9~dB, although the input ripple level was 0.2~dB, corresponding to an ideal return loss of 13.5~dB.
These nearly equal-ripple responses are consistent with the Chebyshev filter design.
No higher-order harmonics are observed up to 350~GHz.
Their design dimensions are listed in Table \ref{tab:bpf_params}.
Here, the symmetry relations $c_{\mathrm{a}(n,n+1)} = c_{\mathrm{c}(N-n,N-n+1)}$ and $c_{\mathrm{b}(n,n+1)} = c_{\mathrm{b}(N-n,N-n+1)} \ (n=0, \dots, N)$ are applied, except for $c_{\mathrm{a}45}$ and $c_{\mathrm{c}45}$ due to the asymmetry of the IDC fingers.
The size of each individual element does not exceed $\sim$ 70~$\mu$m, which corresponds to a quarter of the guided wavelength at 350~GHz. 

\begin{table*}[tb]
        \caption{BPF dimension parameters in $\mu\mathrm{m}$}
        \label{tab:bpf_params}
        \centering
        \begin{tabular}{cc|ccccccc|c} \hline
                 & & & & $c_{\mathrm{a}01}$ & $c_{\mathrm{a}12}$ & $c_{\mathrm{a}23}$ & $c_{\mathrm{a}34}$ & $c_{\mathrm{a}45}$ & \\
                $f_0$ & $BW$ & $w+2s$ & $l_{\mathrm{r}n} = l_\mathrm{r}$ & $c_{\mathrm{b}01}$ & $c_{\mathrm{b}12}$ & $c_{\mathrm{b}23}$ & $c_{\mathrm{b}34}$ & $c_{\mathrm{b}45}$ & Total \\
                 & & & & $c_{\mathrm{c}01}$ & $c_{\mathrm{c}12}$ & $c_{\mathrm{c}23}$ & $c_{\mathrm{c}34}$ & $c_{\mathrm{c}45}$ & \\\hline
                 & & & & 7.3 & 48.1 & 38.4 & 37.2 & 37.0 & \\
                150~GHz & 45~GHz & 52 & 64.7 & IDC 25.6 & IDC 18.9 & IDC 9.5 & IDC 8.2 & IDC 8.0 & 1168.8 \\
                 & & & & 4.2 & 18.7 & 33.5 & 35.5 & 36.0 & \\\hline
                 & & & & 8.5 & 43.6 & 34.0 & 33.6 & 33.5 & \\
                220~GHz & 40~GHz & 28 & 48.8 & IDC 25.6 & IDC 15.8 & IDC 7.8 & IDC 6.8 & IDC 6.6 & 993.4 \\
                 & & & & 4.1 & 21.9 & 31.2 & 32.3 & 32.5 & \\\hline
                 & & & & 8.4 & 35.2 & 27.5 & 26.9 & 26.8 & \\
                270~GHz & 40~GHz & 28 & 40.5 & IDC 19.5 & IDC 8.5 & gap 4.1 & gap 4.6 & gap 4.7 & 807.7 \\
                 & & & & 4.1 & 20.7 & 26.5 & 26.7 & 26.8 & \\\hline
        \end{tabular}
\end{table*}

\subsection{Triplexer}
\begin{figure}[tb]
        \centering
        \subfloat[]{\includegraphics[width=0.6\linewidth]{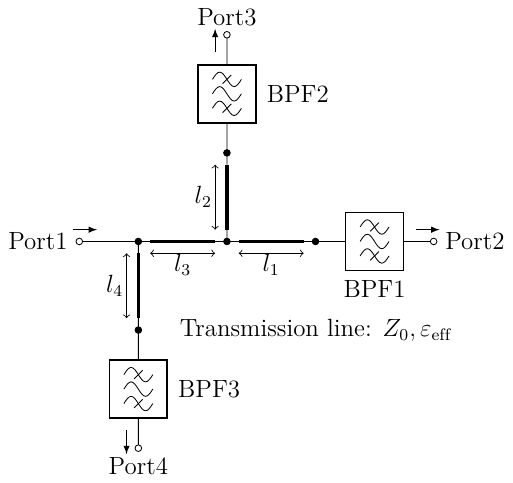}
        \label{subfig:circ_triplexer}}\\
        \subfloat[]{\includegraphics[width=\linewidth]{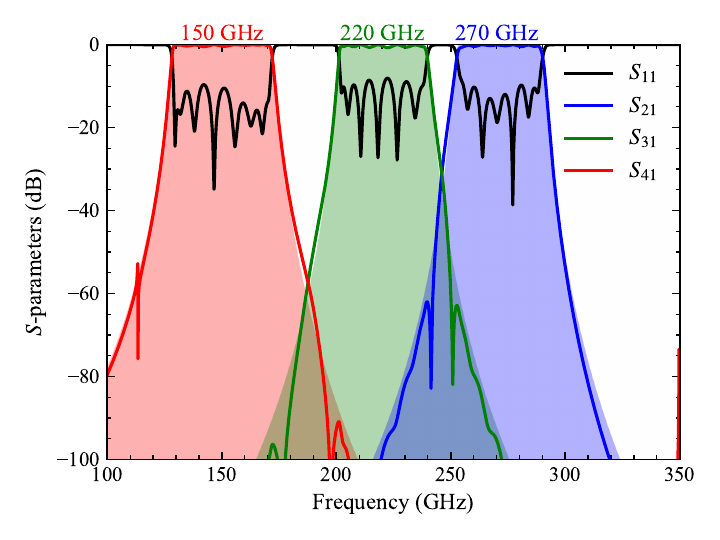}
        \label{subfig:triplexer_dB}}
        \caption{(a) Layout of the manifold multiplexer for the triplexer.
        (b) Simulated frequency response of the triplexer compared with the individual BPFs.
        The red, green, and blue curves correspond to 150, 220, and 270~GHz BPFs, respectively.
        The solid lines represent the $S$-parameters of the triplexer, while the shaded areas indicate the transmittance of the individual BPFs.
        In the network shown in (a), the 270, 220, and 150~GHz BPFs are inserted at BPF1, BPF2, and BPF3, respectively.
        The transmission lines have the following parameters: $Z_0 =$ 89.64~$\Omega$, $\varepsilon_\mathrm{eff} =$ 10.46, 
        $l_1 =$ 150~$\mu$m, $l_2 =$ 120~$\mu$m, $l_3 =$ 130~$\mu$m, and $l_4 =$ 130~$\mu$m.}
        \label{fig:triplexer}
\end{figure}

A millimeter-wave triplexer was constructed by connecting three BPFs in parallel.
We selected the manifold multiplexer topology \cite{Cameron2007} because of its simple structure for a small number of wideband filters.
The circuit and simulated frequency response of the triplexer are shown in Fig.~\ref{fig:triplexer}.
The lengths of the transmission lines connecting the filters were optimized to suppress interference between them.
The frequency response of the triplexer was calculated as that of a synthesized 4-port network, using scattering matrices of BPFs, 90~$\Omega$ transmission lines, and ideal junctions.
While the triplexer response shows a slight shift in passbands compared to the individual BPFs, a low band-to-band crosstalk of less than $-$30~dB was achieved.
The physical size of the triplexer shown in Fig.~\ref{fig:triplexer}(a) is 2.4~mm $\times$ 1.1~mm, corresponding to 3.9$\lambda_\mathrm{g}$ $\times$ 1.8$\lambda_\mathrm{g}$, where $\lambda_\mathrm{g}$ is the guided wavelength at the lowest band of the triplexer.
A comparison of the sizes of our triplexer with those reported in previous works is summarized in Table~\ref{tab:triplexer_literature}.
The relatively large size reflects the high filter order and the feature of quasi-lumped element design.

\begin{table}[tb]
\centering
\begin{threeparttable}[tb]
        \caption{Comparison of millimeter-wave planar triplexers}
        \label{tab:triplexer_literature}
        \centering
        \begin{tabular}{ccccc} \hline
                \multirow{2}{*}{Reference} & \multirow{2}{*}{Structure} & Bands & \multirow{2}{*}{Order} & Size \\
                 & & (GHz) & & ($\lambda_\mathrm{g}^2$) \\\hline
                \cite{Suzuki2012} & Microstrip distributed & 95/150/220 & 3/3/3 & 0.7$\times$0.7\tnote{*} \\
                \cite{thesis_Suzuki2013} & Microstrip lumped & 95/150/220 & 3/3/3 & 1.1$\times$0.3\tnote{*} \\
                \textbf{This work} & \textbf{CPW quasi-lumped} & \textbf{150/220/270} & \textbf{8/8/8} & \textbf{3.9}$\mathbf{\times}$\textbf{1.8} \\\hline
        \end{tabular}
        \begin{tablenotes}
            \item[*] Visual estimates from the figures in the references
        \end{tablenotes}
\end{threeparttable}
\end{table}

\section{Demonstration Using a Scaled Model}
To validate the BPF designed using our proposed method, we designed, fabricated, and measured a scaled model (Fig.~\ref{fig:scaled_model}).

\subsection{Design}
The original filter was designed as an 8th-order Chebyshev BPF fabricated with a Nb film, optimized for $f_0 =$ 150~GHz and $BW =$ 40~GHz, and matched to 50~$\Omega$ ports.
The filter was then scaled by a factor of 20 to create a 7.5~GHz BPF for 2-port measurements below 20~GHz.
The chip size is 40~mm in length and 20~mm in width.
Together with the BPF, a CPW line (center strip width $w =$ 20~$\mu$m, gap width $s =$ 10~$\mu$m) was measured as a transmission reference.
To mitigate interference from substrate mode resonances without using via holes, a $\beta$-phase Ta layer (50~nm thickness, sheet resistance $R_\mathrm{s} =$ 36~$\Omega/\square$ at 4~K) was used as a lossy ground plane (Fig.~\ref{fig:scaled_model}(a)).
The $\beta$-Ta lossy ground was placed at a distance 2.5 times the BPF width from the center axis to ensure proper CPW mode propagation.
Indeed, our preliminary experiment using a lossless ground instead of a lossy ground showed multiple resonances that disturb the intrinsic bandpass response.
Another preliminary investigation revealed that, in the absence of backside metallization, a small gap on the order of 10~$\mu$m between the chip and its holder can induce a non-negligible shift in the cutoff frequency, deviating from the designed bandpass.
To avoid such a passband shift by an unintended gap, a backside Nb layer thicker than the magnetic penetration depth was introduced.

\subsection{Fabrication}
\begin{figure}[tb]
        \centering
        \subfloat[]{\includegraphics[width=0.5\linewidth]{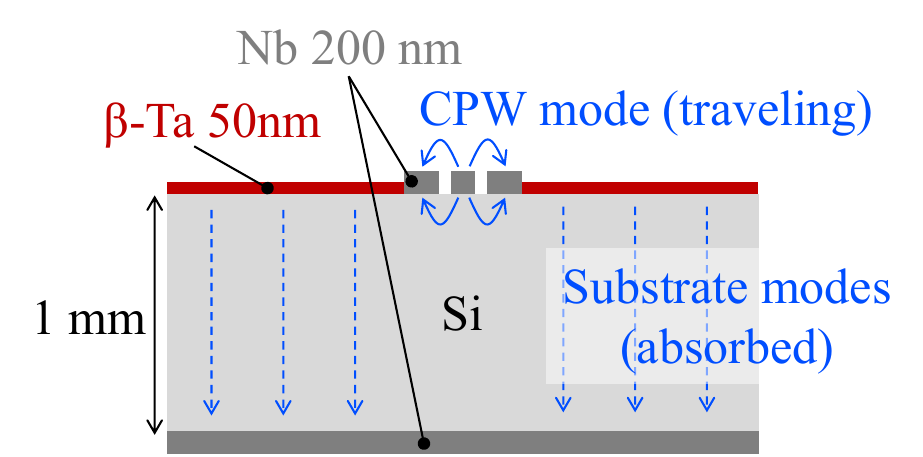}
        \label{subfig:scaled_model_layer}}
        \hfil
        \subfloat[]{\includegraphics[width=0.4\linewidth]{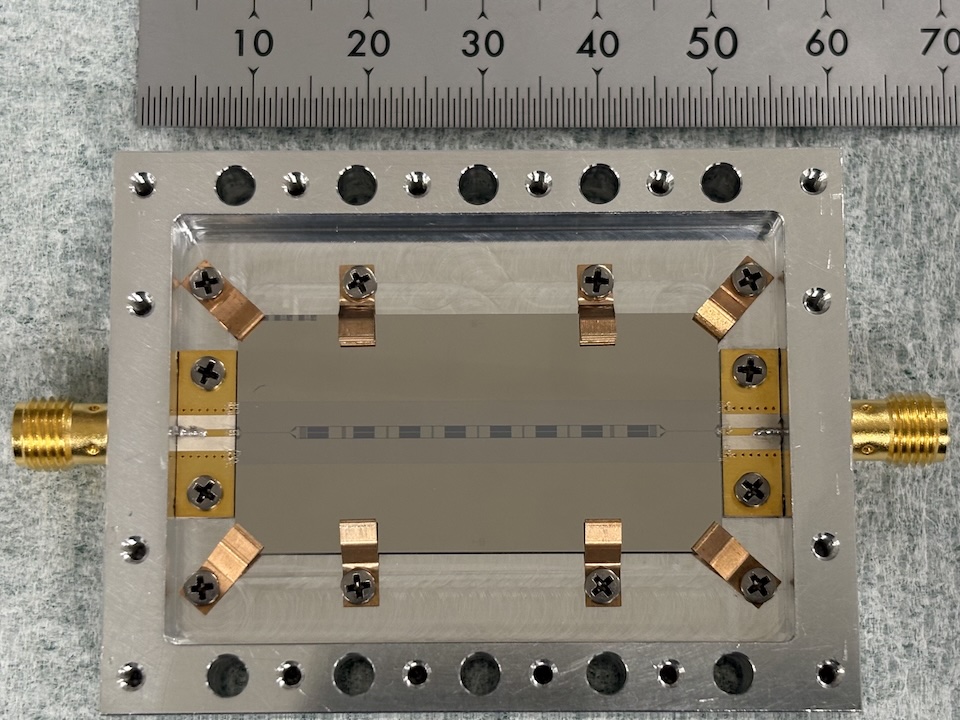}
        \label{subfig:scaled_model_chip}}\\
        \subfloat[]{\includegraphics[width=0.98\linewidth]{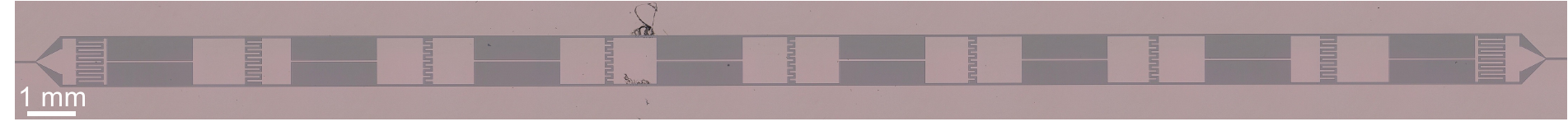}
        \label{subfig:scaled_model_zoom}}
        \caption{(a) Cross-sectional schematic of the scaled model chip.
        (b) Photograph of the scaled model BPF chip mounted on the holder.
        (c) Microscopic image of the BPF structure.}
        \label{fig:scaled_model}
\end{figure}

The fabrication of the BPF chip was carried out at RIKEN.
The substrate was a high resistivity silicon wafer ($>$ 10~k$\Omega\cdot$cm) with a thickness of 1~mm.
After spin-coating photoresist, the $\beta$-Ta pattern was defined by a maskless lithography system (DL-1000SG/RWC, Nano System Solutions, Inc.) and developed.
Then a 50~nm layer of $\beta$-Ta was deposited by DC magnetron sputtering, and patterned via liftoff.
The same liftoff process was repeated to fabricate a 200~nm Nb pattern.
After resist protection on the front side, a 200~nm Nb plane was deposited on the back side of the substrate.
Finally, additional CF$_4$/O$_2$ inductively coupled plasma (ICP) etching was conducted on the frontside Nb pattern to remove fragments which unintentionally survived the liftoff process, as such fragments would short-circuit the center strip and ground plane and severely degrade the bandpass response.
While this additional process leads to over-etching of substrate silicon up to $\sim$ 460~nm, it is expected to have little impact on the performance of the scaled model.
Finally, the wafer was diced into a size of 40~mm $\times$ 20~mm, and each chip was mounted on a holder and wire-bonded at input and output ports.
The fabricated BPF chip is shown in Fig.~\ref{fig:scaled_model}(b)(c).

\subsection{Measurements and Results}
The scattering parameters of the BPF and the reference line were measured at the 4~K stage of a cryostat using a calibrated vector network analyzer (VNA; E8362C, Agilent Technologies).
The VNA calibration was conducted outside the cryostat, and thus the signal paths inside the cryostat were not calibrated.
Although the BPF and the reference line were routed through different signal paths inside the cryostat, the wiring configurations were made nearly identical using reproduced sets of coaxial cables and transitions.
To remove the loss contributions in the signal path outside the BPF, we normalized the transmittance of the BPF by that of the reference line and defined it as the relative transmittance.
The reference line was assumed to have smooth and nearly lossless transmission characteristics.
This assumption is supported by electromagnetic simulations predicting that the reference line exhibits a transmission loss of less than 2\% below 10~GHz, which is smaller than the ripple level of the filter.
The relative transmittance was used for the subsequent analysis of the BPF performance.

\begin{figure}[tb]
        \centering
        \includegraphics[width=\linewidth]{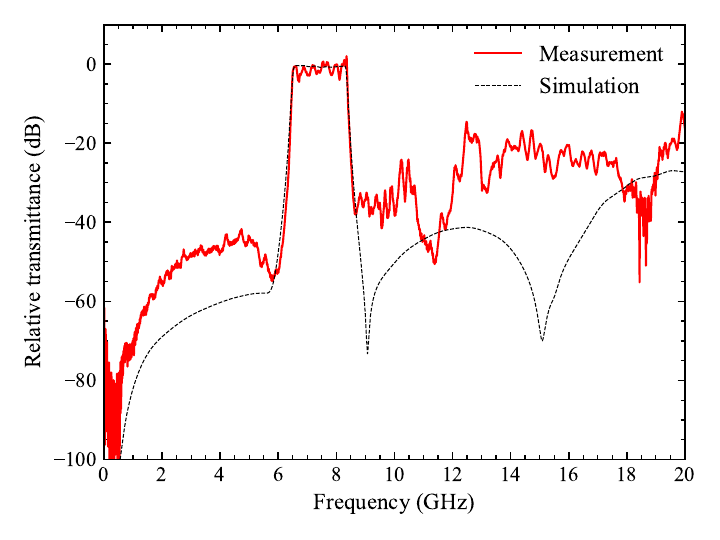}
        \caption{Measured relative transmittance of the scaled model BPF in comparison with simulation.}
        \label{fig:meas_norm}
\end{figure}

Fig.~\ref{fig:meas_norm} shows the relative transmittance compared with simulation by Ansys HFSS \cite{HFSS_webpage}.
The bandpass response over 6.5--8.4~GHz and the steep cutoff are in good agreement with the design.
We note that this passband is slightly narrower than that of the ideal 20$\times$ scaled design, because the wafer thickness (1~mm) is only 5$\times$ the original (200~$\mu$m) rather than the ideal 20$\times$, as confirmed by simulations.
Thanks to the $\beta$-Ta lossy ground, no significant resonances due to substrate modes were observed.
The ripple in the passband is larger than expected, which may be caused by impedance mismatches or reflections at discontinuities such as wire bonds or connectors.
The out-of-band transmittance is also higher compared to the simulation, the origin of which is unclear.
A possible explanation is the uncertainty in electromagnetic modeling around the ports, which may underestimate the coupling to the substrate modes particularly at frequencies above the passband.
This uncertainty does not affect the cutoff frequencies of the BPF.

\section{Conclusion}
We designed quasi-lumped element BPFs for millimeter wavelengths, using CPW structures and compatible with a 2~$\mu$m process rule.
Our efficient design method opens a way to optimize high-order quasi-lumped element BPFs with reasonable computational costs.
We showed the feasibility of a triplexer design by interconnecting the 150, 220, and 270~GHz BPFs optimized using the proposed method.
Moreover, an 8th-order Chebyshev BPF was experimentally demonstrated using a scaled model at 7.5~GHz.
The measurement results show a bandpass response with a steep cutoff as predicted by simulation.
This CPW-based BPF design will simplify the fabrication process and contribute to the development of large pixel arrays for multi-band millimeter and submillimeter imaging cameras in astronomical applications.

\section*{Acknowledgment}
This study was carried out in cooperation with the Advanced Technology Center of the National Astronomical Observatory of Japan (NAOJ).
T.T. was supported by the MEXT Leading Initiative for Excellent Young Researchers (Grant No. JPMXS0320200188).
This work was supported in part by RIKEN Special Postdoctoral Researcher Program,
JSPS KAKENHI Grant Numbers JP23H00121, JP23K25879, JP23K25905, JP23K20035, JP24H00004, and JP24K22911, 
the Murata Science and Education Foundation, the Nakajima Foundation, and the Sumitomo Foundation (Basic Science Research Grant No. 2200541).
A part of this work has been done at Nanoscience Joint Laboratory, RIKEN.
The authors would like to thank Dr. Chiko Otani for kindly offering access to fabrication facilities in RIKEN.

\bibliographystyle{IEEEtran}
\bibliography{IEEEabrv,biblist}

\end{document}